\documentclass[aps,prl,showpacs,amsmath,amssymb,twocolumn]{revtex4}
\usepackage{graphicx}
\usepackage{dcolumn}
\begin{document}
\title{Ferromagnetic Annulus And Superconducting Vortices}
\author{M. Amin Kayali}
\email{amin@rainbow.physics.tamu.edu}
\affiliation{Department of Physics, Texas A \& M University\\College Station, Texas 77843-4242, USA}

\begin{abstract}
\noindent
The screening currents induced in a superconducting film by a magnetic annulus whose magnetization is perpendicular to the superconductor are calculated. We show that close to the superconductor transition temperature $T_c$  particular values of the magnetization and radii of the annulus make the creation of superconducting vortices energy favorable. We also show that the magnetic annulus offers an alternative tool for vortex pinning in the superconductor. Statistical mechanical properties of this system are discussed.\\

\end{abstract}
\maketitle

Recent studies on magneto-superconducting systems have shown that such systems can be very interesting from theoretical and experimental point of views. Not only are these systems  important for the magnetic storage systems industry, but also they hold great promises in superconducting technology. It is known that the usefulness of any superconductor is determined by the critical current it can pass. Type I superconductors have small critical currents; therefore, the focus is more on Type II superconductors. In type II superconductor vortices appear and lead to higher values of critical current. Unfortunately, the thermal motion of vortices in the superconductor eventually destroy superconductivity(SC). Hence it is crucial to stop the motion of the vortices. The motion of the vortices can be stopped by using magnetic pins or lattic defects. However, magnetic pins proved to be better tools for pinning the vortices, see \cite{ketter}$-$\cite{sasik}.\\

In \cite{marmor} Marmorkos {\it{et.al.}} considered the problem of a magnetic dot embedded into a superconductor. The dot magnetization is directed along the perpendicular direction to the superconductor. They solved the non-linear Ginzburg-Landau equation numerically using appropriate boundary conditions. They found that for some values of the magnetization per unit area of the dot $m$, the dot will couple to a vortex state with $n$ flux quanta .R. Sasik and T. Hwa in \cite{sasik} also studied the vortex pinning by an array of magnetic dots. They concluded that a dot can pin a vortex with multiple flux quanta. However, in their study they ignored the geometry of the dot by considering it a dipole like pin. The geometry of the dot was considered in \cite{ours}, where it was found that not only is the dot a good pin, but it can also spontaneously creat vortices in the superconductor forming a Dot-Vortex bound state.\\
   
When a ferromagnetic (FM) and a superconducting systems are brought together, the order parameters for both systems suppress each other (the proximity effect) in the space between the two systems, and weaken it outside. This fact leads to the loss of both phenomena. However, to avoid the proximity effect and the spin diffussion, Pokrovsky and Lyuksyutov \cite{pok1dot,pk2dot} proposed to seperate the two systems by an infinitely thin layer of insulating oxide. The insulator thickness should be of the order of the coherence length $\xi$ of the superconductor which is much smaller than the superconductor's magnetic penetration depth $\lambda$.\\

In this article we consider the interaction between a ferromagnetic annulus and a superconducting thin film. The annulus lies in the $xy$-plane and its magnetization is directed along the positive $z$ axis. The superconductor is positioned on the top of the annulus with no electrical contact between the ferromagnet and the superconductor. The annulus generates a magnetic field which penetrates into the superconductor and consequently screening currents appear in the superconductor. The screening currents generate magnetic fields which interact with the ferromagnet. We studied the system in the case when vortices in the superconductor appeared as a result of an external magnetic field. Also, we considered the creation of the Annulus-Vortex bound states in the absence of the external magnetic field. \\ 

This article is organized as follows: In the next section we present the calculations for the magnetic field from the annulus and the screening currents and their asymptotic behavior. Section three will present the calculation for  the energy for the coupled annulus-vortex system and discuss its implications such as pinning. Then we conclude with some remarks on the statistical mechanical properties of this system.

\section{The FA-SC System}
Let us consider a ferromagnetic annulus in the $xy$-plane whose thickness is $d_m\ll \lambda$. The inner radius of the FA is $a$ and outer one is $b$. The magnetization of the annulus is directed along the positive $z$-direction. On the top of the FA we put a superconducting film with thickness $d_s \ll \lambda$. The two systems are separated by a thin layer of insulator with thickness $D$ which is of the order of the coherence length $\xi$.\\

\noindent
The magnetization of the annulus can be written as follows:
\[ {\bf M}(\rho,\varphi,z)=\left\{ \begin{array}{rll}
			0 \hspace{2.0cm} \rho < a; \\
			m_0\delta(z) \hat{z}\hspace{0.5cm} a\leq \rho \leq b;\\
			0 \hspace{2.0cm} \rho > b.
			        \end{array} \right.
\]
\noindent
where $m_0$ is the magnetization per unit area.\\

In the presence of the superconductor, the London-Maxwell equation can be expressed as follows:

\begin{equation}
{\bf \nabla}\times{\bf \nabla}\times{\bf A} = -\frac{1}{\lambda} {\bf A}\delta(z) +\frac{n\phi_0}{2\pi\lambda}{\bf \nabla}\phi\delta(z) +4\pi {\bf \nabla}\times{\bf M}
\end{equation}

\noindent
where $n$ is the vorticity of the superconducting vortex, $\phi_0=hc/2e$ is the flux quantum and $\lambda=\frac{\lambda_L^2}{d_s}$ is the effective magnetic penetration depth. The vector field ${\bf A}$ is the total vector field contributed by the ferromagnet ${\bf A}_m$ and the superconductor ${\bf A}_s$. The system has azimuthal symmetry, then on using the gauge ${\bf \nabla}.{\bf A}=0$, eq.(1) becomes:

\begin{equation} 
-{\bf \nabla}^2 {\bf A} = -\frac{1}{\lambda} {\bf A}\delta(z) +\frac{n\phi_0}{2\pi\lambda}{\bf \nabla}\phi\delta(z) +4\pi {\bf \nabla}\times {\bf M}
\end{equation}

\noindent
Using the superposition principle we can separate eq.(2) into two sets of equations which can be expressed as follows:

\begin{equation}
-{\bf \nabla}^2 {\bf A}_s +\frac{1}{\lambda} {\bf A}_s\delta(z)=\frac{n\phi_0}{2\pi \lambda} \frac{\hat{z}\times (\vec{\rho} -\vec{\rho_0})}{|\vec{\rho} -\vec{\rho_0}|^2}\delta(z)
\end{equation}

\begin{equation}
-{\bf \nabla}^2 {\bf A}_m +\frac{1}{\lambda} {\bf A}_m \delta(z)=4\pi {\bf \nabla}\times {\bf M}
\end{equation}

\noindent
${\bf A}_s$ is the vector potential produced by a superconducting vortex centered at $\rho_0$, which can be written as follows:

\begin{equation}
{\bf A}_s =\frac{n\phi_0}{2\pi}\frac{\hat{z}\times(\vec{\rho} -\vec{\rho_0})}{|\vec{\rho} -\vec{\rho_0}|} \int_0^\infty \frac{J_1(q|\vec{\rho} -\vec{\rho_0}|)}{1 +2\lambda q}e^{-q|z|}dq
\end{equation}

\noindent
where $J_n(q\rho)$ is the $n^{th}$ order Bessel function. The Fourier transform of ${\bf A}_m$ is:

\begin{equation}
{\bf A}_m(\rho, \varphi,z) =\frac{1}{(2\pi)^3}\int {\bf A}_m({\bf K})e^{-\imath {\bf K}.{\bf r}} d^3 K
\end{equation}
\noindent
where ${\bf K}=({\bf q},k_z)$ is the wave vector. Then the London-Maxwell equation for ${\bf A}_m$ becomes:

\begin{equation}
{\bf K}^2 {\bf A}_m ({\bf K}) +\frac{1}{\lambda} {\bf a}_m({\bf q})=8\pi^2 \imath m_0 \left(bJ_1(qb)-aJ_1(qa)\right)\hat{{\bf \varphi}}_q
\end{equation}
\noindent
where
\begin{equation}
{\bf a}_m ({\bf q})=\frac{1}{2\pi}\int_{-\infty}^\infty {\bf A}_m ({\bf K})dk_z
\end{equation}

\noindent
with a bit of algebra we find:

\begin{equation}
{\bf A}_m =4\pi m_0 \hat{\varphi} \int_0^\infty q\frac{\left[bJ_1(qb) -aJ_1(qa)\right]J_1(q\rho)}{1 +2q}e^{-q|z|} dq
\end{equation}
\noindent
Using ${\bf B}={\bf \nabla}\times {\bf A}$ we calculate the $z$-component of the magnetic field due the FA system:

\begin{equation}
B_m^\rho =-\frac{4\pi m_0}{\lambda} \int_0^\infty  \frac{q^2 \left[bJ_1(qb) -aJ_1(qa)\right]J_1(q\rho)}{1 +2q}e^{-q|z|} dq
\end{equation}

\begin{equation}
B_m^z=\frac{4\pi m_0}{\lambda} \int_0^\infty  \frac{q^2 \left[bJ_1(qb) -aJ_1(qa)\right]J_0(q\rho)}{1 +2q}e^{-q|z|} dq 
\end{equation}
\noindent 
where in eqs.$(9-11)$ we rescaled the variables such that $q\rightarrow \lambda q$, $(\rho,z)\rightarrow (\frac{\rho}{\lambda},\frac{z}{\lambda})$, and $(a,b)\rightarrow (\frac{a}{\lambda},\frac{b}{\lambda})$.\\

The total $z$-component of the magnetic field on the film is the sum of the one from the magnetic annulus and the SV. The $z$-component of the magnetic field from the SV behaves like $\frac{1}{\rho^2}$ at distances smaller than $\lambda$,  and like $\frac{1}{\rho^3}$ at large distances. The behaviour of the z-component of the annulus magnetic field is shown in Fig.(1).
\begin{figure}[h]
 \begin{center}
 \includegraphics[angle=270,width=3.0in,totalheight=1.9in]{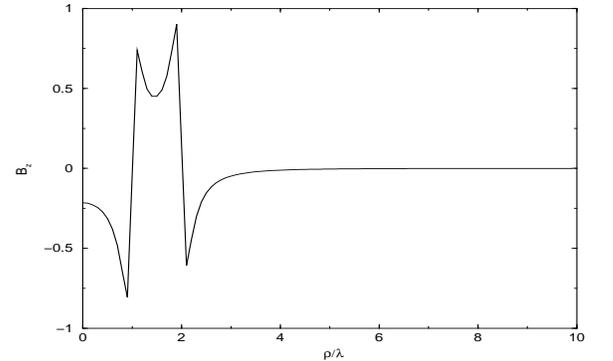}
  \caption{The $z$-component of the annulus magnetic field on the surface of the superconductor. The annulus has a radii $(a=\lambda,b=2\lambda)$ and the field is measured in units of $\frac{4\pi m_0}{\lambda}$.}
 \end{center}
\end{figure}

\section{The Energy of the FA-SV System}
We calculate the energy for the FA-SV system assuming a superconducting vortex with vorticity $n$ is centered at $\rho_0$. The total energy of this system is:

\begin{equation}
E=\int \left[ \frac{{\bf B}^2}{8\pi} +\frac{mn_s {\bf v}_s^2}{2} - {\bf M}.{\bf B}\right] dV
\end{equation}

\noindent
Since the two systems are planar, then we can use the argument in \cite{ours} to rewrite eq.(12) as follows:

\begin{equation}
E=\int \left[ \frac{n_s \hbar^2}{2m}\left( {\bf \nabla}\varphi\right)^2 -\frac{n_s\hbar e}{2mc} {\bf \nabla}\varphi.{\bf a} -\frac{1}{2} {\bf m}.{\bf b}\right] d^2x
\end{equation}

\noindent
using eqs.(5,9,11) in (13) we find:

\begin{eqnarray}
E &=& n^2 \epsilon_0 ln(\frac{\lambda}{\xi}) 
-n\epsilon_m \int_0^\infty \frac{\left(bJ_1(qb)-aJ_1(qa)\right)J_0(q\rho_0)}
{1+ 2q} dq \nonumber \\  
& &-4\pi^2 m_0^2 \int_0^\infty \frac{\left[bJ_1(qb)-aJ_1(qa)\right]^2}{1+2q} dq
\end{eqnarray}

\noindent
where $\epsilon_0=\frac{\phi_0^2}{16\pi^2 \lambda}$ is the energy scale for the vortex energy. and $\epsilon_m=m_0 \phi_0$ is the energy scale for the magnetic interaction between the annulus and the vortex. The $\epsilon_m$ can easily be made larger than $\epsilon_0$ especially near the superconducting transition temperature when $\lambda$ becomes infinite. The ratio $\delta_m=\frac{\epsilon_m}{\epsilon_0}$ can be written as follows:
\begin{equation}
\delta_m=gS\frac{2n_m d_m}{n_s d_s}
\end{equation}
\noindent
where $g$ is the Lande factor, $S$ is the magnet elementary spin, $n_m$ is the magnet atomic density and $d_s, d_m$ are the thicknesses of the superconductor and the magnet respectively.

\noindent 
We integrated eq.(11) numerically using different integration techniques and we plot the energy of this system as a function of $\rho_0$. We found that the energy has a minimum when $a<\rho_0<b$ as shown in Fig.(2).\\

\begin{figure}[h]
 \centering
 \includegraphics[angle=270,width=3in,totalheight=1.9in]{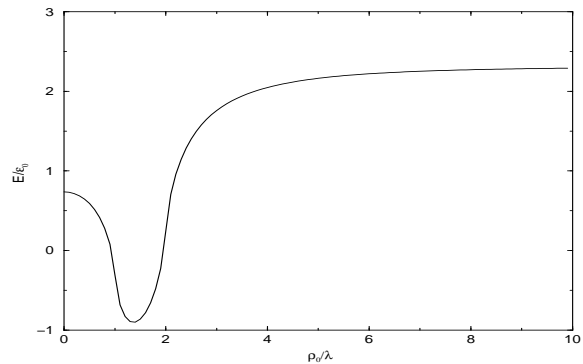}
 \caption{The energy in units of $\epsilon_0$ as a function of 
$\frac{\rho_0}{\lambda}$ for the case when $n=1, \frac{\lambda}{\xi}=10, a=\lambda,b=2\lambda$ and $\delta_m=10$.}
 \label{fig:energy}
\end{figure}

\noindent
This shows that the vortex will not be centered at the origin as one may naively expect from the symmetry of the system. It rather prefers to move its center over the magnetized region of the annulus to minimize the energy of the system. Moreover, since the annulus has an azimuthal symmetry we find a continuous set of choices for the vortex center along a circle of radius $\rho_0$.

The energy plot shows that even in the absence of the external magnetic field the annulus can create vortices in the superconductor if the magnetic field produced by the annulus exceeds the matching field $H_{\phi}$ of the superconductor. This can be controlled by changing the thickness of the magnetic annulus or $n_s$. Let us call the second integral in eq.(14) $\Gamma_{ab\rho_0}$, and minimize the energy of the system with respect to $n$ to find the maximum magnetic flux charge the annulus can pin. \\

\noindent
To this end we recognize two cases. The first case is when vortices are created by an external magnetic field. The second one is when the external field is absent. These two cases were recognized earlier in \cite{sasik}. However, in the geometry we study here we notice that if the vortices appeared due to an external magnetic field then the annulus will attract the vortices and trap their centers over it since the energy is minimal in that case. The maximum charge $n_c^\prime$ the annulus can pin in this case can be calculated. For this purpose we assume that the total number of vortices produced in the superconductor due to the  external magnetic field is $N$ and that the annulus has trapped a vortex of charge $n$. Then, if we assume that the rest of the vortices are singly quantized, we find after minimization that the maximum charge the annulus would trap in the presence of the external field is:

\begin{eqnarray}
n_c^\prime= Int\left[\frac{1}{2} +\delta_m \frac{\Gamma_{ab\rho_0}}{2ln(\frac{\lambda}{\xi})}\right]
\end{eqnarray}
\noindent
where $Int[P]$ stands for the closest integer to $P$.

Moreover, in the absence of the external magnetic field the annulus simultaneously creates $n_c$ vortices in the superconductor and couples to them provided that the annulus magnetic field $B_m\geq n_c B_\phi$ where $B_\phi$ is the matching field. The number of the magnetic flux quanta simultaneously generated by the annulus is: \\

\begin{equation}
n_c =\left[ \frac{\delta_m \Gamma_{ab\rho_0}}{2 ln(\frac{\lambda}{\xi})} \right]
\end{equation}

\noindent
In conclusion, we found that the magnetic annulus offers an alternative pinning center for vortices in a superconductor. When vortices exist in the superconductor due to an external magnetic field the annulus attracts vortices and confines some of them under it to minimize the energy of the system. In the absence of the external magnetic field, the ferromagnetic annulus can create vortices in the superconductor and form a bound system of FA-SV, provided that the $z$-component of the annulus magnetic field exceedes the matching field for creating a vortex. Our calculations showed that the vortex trapped or created under the annulus may have one or more flux quanta depending on the ratio $\delta_m$ which can be controlled easily either by changing the thicknesses of the films or by decreasing the density of the superconducting carriers. In the superconducting phase near the transition temperature the ratio $\delta_m$ becomes very large, creating a giant vortex above the annulus.  The center of the trapped vortex can be any point on a circular manifold which lies above the annulus and its radius is $\rho_0$, where $a< \rho_0 <b$. This degeneracy of the vortex center location can be lifted by squashing the annulus. A study of the squashed magnetic dots and magneic annulus will be published in a separate article.\\
\noindent\\
This work was partly supported by DE-FG03-96ER45598, and NSF DMR-00-72115.I would like to thank V. L. Pokrovsky, I. F. Lyuksyutov and D. Naugle for the stimulating discussion and valuable suggestions.


\begin{references}
\bibitem{ketter} D. J. Morgan and J. B. Ketterson, Phys. Rev. Lett. {\bf 80}, 3614(1998).

\bibitem{bula} L. N. Bulaevskii, A. I. Buzdin, M. L. Kulic and S. V. Panyukov, Adv.Phys. {\bf 34}, 175 (1985).

\bibitem{schull} J. I. Martin, M. Velez, J. Nogues, and I. K. Schuller  Phys. Rev. Lett. {\bf 79}, 1929, (1997). 

\bibitem{pok1dot} I. F. Lyuksyutov and V. L. Pokrovsky Phys. Rev. Lett. {\bf 81}, 2344 (1998).   

\bibitem{pk2dot} I. F. Lyuksyutov and V. L. Pokrovsky, {\it{Magnetism Controlled Vortex Matter}} cond-mat/9903312.

\bibitem{ours} S. Erdin, A. F. Kayali, I. F. Lyuksyutov and V. L. Pokrovsky, {\it{Interaction of Mesoscopic Magnetic Textures with Superconductors}} cond-mat/0007016.
\
\bibitem{naugle} I. F. Lyuksyutov and D. G. Naugle, Modern Phys. Lett. B {\bf 13}, 491 (1999).

\bibitem{marmor} I. K. Marmorkos, A. Matulis and F. M. Peeters Phys. Rev. B {\bf 53}, 2677 (1996).

\bibitem{sasik} R. Sasik, T. Hwa, {\it{Enhanced Pinning of Vortices in Thin Film Superconductors by Magnetic Dot Arrays}} cond-mat/0003462.

\end{references}
\end{document}